\documentclass[preprintnumbers,3p,12pt]{elsarticle}

\usepackage{amsmath,amssymb}
\usepackage{mathtools}
\usepackage{bm}
\usepackage{graphicx}
\usepackage{MnSymbol}
\usepackage{url}
\usepackage{hyperref}

\newcommand{\msbar}{\overline{\rm MS}}

\newcommand{\bea}{\begin{eqnarray}}
\newcommand{\eea}{\end{eqnarray}}
\newcommand{\simgt}{\hbox{ \raise3pt\hbox to 0pt{$>$}\raise-3pt\hbox{$\sim$} }}
\newcommand{\simlt}{\hbox{ \raise3pt\hbox to 0pt{$<$}\raise-3pt\hbox{$\sim$} }}

\newcommand{\be}{\begin{equation}}
\newcommand{\ee}{\end{equation}}

\journal{Physics Letters B}

\begin{document}

\begin{flushright}
    \normalsize TU--1251
\end{flushright}

\begin{frontmatter}

\title{Two-loop $O(\epsilon)$ term of $1/(mr^2)$ heavy quarkonium potential}
\author{Yukinari~Sumino
\vspace*{3mm}
}

\address{
Department of Physics, Tohoku University,
Sendai, 980--8578 Japan
}%

\begin{abstract}
\small
As a straightforward application of the recently calculated two-loop heavy quarkonium Hamiltonian,
we evaluate the two-loop $O(\epsilon)$ term of the $1/(mr^2)$ heavy quarkonium potential. 
Compared to a previous
 calculation we find a small difference in the coefficient of the maximally non-abelian color factor $C_F C_A^2 $.
 We further examine this coefficient carefully.
\end{abstract}

\end{frontmatter}


The heavy quarkonium system, such as bottomonium and charmonium states, has served not only as a valuable laboratory for elucidating QCD dynamics but also as an important tool for precise determination of key parameters in the Standard Model of particle physics, including $m_b$, $m_c$, and $\alpha_s$. 
To support these aims, substantial efforts have been dedicated over the years to improving the theoretical precision of heavy quarkonium observable calculations. 
Recently, progress has been marked by the computation of the two-loop heavy quarkonium Hamiltonian in the non-annihilation channel as a new step towards the next-to-next-to-next-to-next-to-leading order (N$^4$LO) calculations
\cite{Mishima:2024afk}.
The Hamiltonian is defined within the framework of the potential-NRQCD effective field theory 
\cite{Brambilla:2004jw}.


The $O(\epsilon^{-1})$
and $O(\epsilon^0)$ terms of the two-loop $1/(mr^2)$ potential were first computed in ref.~\cite{Kniehl:2001ju} as a part of the N$^3$LO Hamiltonian.
($\epsilon$ is the parameter of the dimensional regularization, where we set the spacetime dimensions as
$D=4-2\epsilon$.)
This result was confirmed in the calculation of the above two-loop Hamiltonian.
As a next step, in this paper we calculate the $O(\epsilon)$ term of the $1/(mr^2)$ potential from the two-loop Hamiltonian. 
This term was first computed in ref.~\cite{Beneke:2014qea} as a necessary ingredient in the computation of the 
partial decay width of the bottomonium vector $1S$ state into a lepton pair $\Gamma(\Upsilon(1S)\to\ell^+\ell^-)$
at N$^3$LO.
Since this $O(\epsilon)$ term of the potential also affects important physical quantities such as the $t\bar{t}$ threshold
cross sections in $e^+e^-$ and hadron collider experiments, we provide an independent calculation.
As it turns out, we find a small difference in the coefficient of the maximally non-abelian color factor $C_F C_A^2 $
of this term.
We further examine this coefficient carefully  by calculations using different methods.

It is straightforward to obtain the $O(\epsilon)$ term of the $1/(mr^2)$ potential from 
the expression for the two-loop Hamiltonian in general dimensions \cite{Mishima:prep}.
In momentum space the potential reads
\bea
&&
V_{1/(mk)}^{\text{(2 loop)}}(k)\biggr|_{{\cal O}(\epsilon)}=
\frac{\pi  C_F \alpha _s(k){}^3}{mk}\times \epsilon
\nonumber\\&&
~~~~~~~~~~~~~~~~~~~~~~
\times
\Biggl[
\left(-\frac{631}{108}-\frac{15 \pi ^2}{16}-\frac{8 \log ^2(2)}{3}+\frac{65 \log
   (2)}{9}\right) C_A C_F
\nonumber\\&&
~~~~~~~~~~~~~~~~~~~~~~~~~
+\left(-\frac{205}{27}-\frac{161 \pi ^2}{72}-\frac{4 \log
   ^2(2)}{3}-\frac{101 \log (2)}{18}\right) C_A^2
\nonumber\\&&
~~~~~~~~~~~~~~~~~~~~~~~~~
   +\left(\frac{115}{108}+\frac{5 \pi ^2}{36}+\frac{49 \log
   (2)}{36}\right) C_A n_l+\left(\frac{17}{54}-\frac{11 \pi
   ^2}{72}-\frac{2 \log (2)}{9}\right) C_F n_l
   \biggr]
   \,.
\nonumber\\&&
\label{OurResult}
\eea
The potential depends on the choice of the operator basis of the Hamiltonian, and our basis agrees with that of refs.~\cite{Kniehl:2001ju,Kniehl:2002br,Beneke:2014qea}.
The first term of $C_A^2$ differs from that of Eq.~(12) of ref.~\cite{Beneke:2014qea}:
we obtain\footnote{
We thank the authors of ref.~\cite{Beneke:2014qea} for confirming this error in their result.
Calculating the potential using method (ii) discussed below facilitated identifying the origin of the error.
} 
$-\frac{205}{27}$ instead of $-\frac{1451}{216}$.
(Numerically the coefficient of $C_A^2$ changes only by a few percent.)
All the other terms agree.
Here, $\alpha_s$ denotes the strong coupling constant in the $\msbar$ scheme for the theory with
$n_l$ massless quark flavors only;
$C_A=N_c$ and $C_F=(N_c^2-1)/(2N_c)$ denote the color factors of the $SU(N_c)$ group.
For completeness we give the two-loop $1/(mr^2)$ potential in general 
dimensions in the Appendix.

In ref.~\cite{Mishima:2024afk}, the two-loop Hamiltonian was calculated by matching the on-shell 
quark-antiquark ($Q\bar{Q}$) scattering amplitudes in full QCD
and pNRQCD.
To check our result, we calculate the $C_FC_A^2$ part of eq.~\eqref{OurResult} in two other methods:
(i) Matching the off-shell amplitudes in full QCD and pNRQCD, and 
(ii) Matching the off-shell amplitudes in NRQCD and pNRQCD.
The method (ii) is close to that of the original calculations \cite{Kniehl:2001ju,Beneke:2014qea}.
Our calculations in (i) and (ii) are carried out in Feynman gauge.

For comparison, let us first briefly explain the calculation by the on-shell matching.
The two-loop on-shell amplitude for 
$Q(\vec{p})+\bar{Q}(-\vec{p})\to Q(\vec{p}+\vec{k})+\bar{Q}(-\vec{p}-\vec{k})$ with 
$|\vec{p}|=|\vec{p}+\vec{k}|$
is calculated in full QCD.
The amplitude is projected to a spinor basis, by which the coefficient of each spinor structure is
expressed by fully relativistic scalar integrals.
The integrals are reduced to master integrals using the integration-by-parts (IBP)
recurrence relations
\cite{Chetyrkin:1981qh}.
Since the on-shell amplitude has poles at $u=\vec{k}^2-4\vec{p}^{\,2}=0$ on unphysical sheets of the complex plane,
the coefficients of the master integrals have poles at $u=0$.
The master integrals are expanded in $1/m$ using combination of the expansion-by-regions (EBR) technique \cite{Beneke:1997zp} and the differential equation satisfied by the master integrals \cite{Gehrmann:1999as}.
We match the amplitude to that of the pNRQCD amplitude order by order in expansions in $\alpha_s$ and
$1/m$.
The $C_A^2C_F$ part of $V_{1/(mk)}^{\text{(2 loop)}}$ is equal to
the $C_A^2C_F$, ${\cal O}(1/m)$ part of the two-loop full QCD
amplitude, because there is no contribution from 
the pNRQCD amplitude except from $V_{1/(mk)}^{\text{(2 loop)}}$.
Hence, at intermediate stages of the computation of the potential,
poles at $u=0$ appear.
Only when we combine all the contributions which
contribute to the $C_A^2C_F$ part of $V_{1/(mk)}^{\text{(2 loop)}}$,
the $p$ dependence drops out and the potential which
depends only on $k$ is obtained.

In contrast, in the case of the off-shell matching,
we set the four momenta of the off-shell initial and final quark and antiquark as
$Q(m+E/2,\vec{p})+\bar{Q}(m+E/2,-\vec{p})\to Q(m+E/2,\vec{p}+\vec{k})+\bar{Q}(m+E/2,-\vec{p}-\vec{k})$.
No pole at $u=0$ appears in the calculations of $V_{1/(mk)}^{\text{(2 loop)}}$.
At the end of the calculation, 
we set $|\vec{p}|^2, |\vec{p}+\vec{k}|^2 \to m E$ to take the on-shell limit, 
corresponding to our choice of the operator basis of the Hamiltonian.
Then the $E$ dependence disappears and the potential which
depends only on $k$ is obtained.

In method (i), starting from the full QCD amplitude, we rewrite the four-component spinors in terms of the 
two-component spinors using the Pauli matrices instead of the $\gamma$ matrices.
Loop integrals are expressed in terms of tensor integrals which are contracted with 
the two-component spinor structures.
We expand the amplitude in $1/m$
using the EBR technique, where only the soft region
contributes to $V_{1/(mk)}^{\text{(2 loop)}}$.
The energy measured from the threshold $E$ is counted as order $\beta^2$, while $\vec{p}$, $\vec{k}$
and loop momenta $l_1,l_2$
are counted as order $\beta$, in this expansion.
The obtained tensor loop integrals in the soft region are expressed in terms of scalar integrals
by form factor decomposition.\footnote{
This is an alternative way to the spinor basis projection used in the on-shell matching for reducing the 
tensor integrals to scalar integrals.
To perform non-trivial cross checks, we adopt independent routes of the calculations as much
as possible.
}

The two-loop scalar integrals take the following form:
\bea
&&
J_{SS}( n_1, \cdots , n_{11} ) =
\int \frac{d^{4-2\epsilon} l_1}{(4\pi)^{4-2\epsilon}} \frac{d^{4-2\epsilon} l_2}{(4\pi)^{4-2\epsilon}} \,
\frac{1}{D_1^{n_1} \cdots D_{11}^{n_{11}}}
\,,
\eea
where the eleven indices $n_1, \cdots , n_{11}$ represent the powers of the propagator denominators,
\bea
&&
\{D_1,\cdots,D_{11}\}
=
\{
l_1\!\cdot\! v,\, l_2\!\cdot\! v,\, (l_1+l_2)\!\cdot\! v,\, l_1^2,\, l_2^2,\, 
\nonumber\\
&&~~~
~~~~~~~~~~~~~~~~~~
(l_1+l_2)^2,\, (l_1+k_1)^2,\, (l_2+k_1)^2,\, (l_1+l_2+k_1)^2,\, l_1\!\cdot\! p_r,\, l_2\!\cdot\! p_r
\}\,.
\eea
The Feynman prescription $+i0$ is suppressed in $D_i$.
The external vectors are given by
$v=(1,\vec{0})$, $k_1=(0,\vec{k})$, $p_r=(0,\vec{p})$ in the c.m.\ frame.
$D_{10}$ and $D_{11}$ appear only in the numerator, i.e., $n_{10},n_{11}\leq 0$.
Hence, each integral is proportional to, or a linear combination of,
$|\vec{p}|^{2n}(\vec{p}\cdot\vec{k})^{n'}|\vec{k}|^{n''-4\epsilon}$ 
for integer $n,n',n''$ and $n,n'\ge 0$.
The amplitude is expressed as a linear sum of $J_{SS}$'s, where the
coefficients of $J_{SS}$'s are given by polynomials of $E,\vec{p},\vec{k}$ contracted
with the spinor structures, apart from a possible
positive integer power of $1/|\vec{k}|$.
By IBP reduction we can express $J_{SS}$'s in terms of three master integrals
$J_{SS}(0, 0, 1, 0, 1, 0, 1, 1, 0, 0, 0)$, $J_{SS}(0, 0, 1, 0, 1, 1, 0, 1, 1, 0, 0)$, $J_{SS}(0, 0, 1, 1, 1, 0, 0, 0, 1, 0, 0)$.
At this stage the coefficients of the spin-dependent structures at ${\cal O}(1/m)$
vanish and only the spin-independent part 
remains.
Off-shell effects are proportional to $E-|\vec{p}|^2/m$ or $E-|\vec{p}+\vec{k}|^2/m$,
such that dependence on $E$ vanishes as we take the limit $|\vec{p}|^2, |\vec{p}+\vec{k}|^2 \to m E$
at the final stage.

In method (ii), we calculate the two-loop off-shell scattering amplitude
using the Feynman rules of NRQCD as given in ref.~\cite{Beneke:2013jia}.
We expand the heavy quark propagators in $1/m$ about the static ones ($1/D_1$, $1/D_2$, $1/D_3$).
Only the vertices at ${\cal O}(m^0)$ and ${\cal O}(1/m)$ are included.
Of the latter vertices only the spin-independent ones contribute at order $1/m$ of the
amplitude, since the amplitude should be symmetric under the exchange of $\vec{p}$
and $\vec{p}'=\vec{p}+\vec{k}$.\footnote{
The only contraction of the spin-dependent vertex 
$[\sigma^i,\sigma^j]/m$ that could survive is
that with $p^i p'^j-p'^i p^j$, which is antisymmetric under the exchange of $\vec{p}$ and $\vec{p}'$.
}
Thus, the $C_FC_A^2$ part of the ${\cal O}(1/m)$ amplitude is expressed by scalar integrals
$J_{SS}$.
After the IBP reduction, they are expressed by the three master integrals. 
The calculation is similar to but simpler than the method (i) since the spin dependence is absent
from the diagram generation.

The final results agree with the $C_FC_A^2$ part of
eq.~\eqref{OurResult} [or eq.~\eqref{PotGeneDim}] in both methods (i) and (ii).
As explained, the methods of the calculations are fairly different, which serve as 
non-trivial cross checks of the result.

\appendix
\section{$1/(mr^2)$ potential for general dimensions}

The two-loop $1/(mr^2)$ potential in momentum space before expansion in $\epsilon$ is
given by
\bea
&&
V_{1/(mk)}^{\text{(2 loop)}}(k) =
\frac{4\pi^3 C_F\alpha _s(k){}^3}{mk} \times 
\left(\frac{e^{\gamma_E}}{4\pi}\right)^{2\epsilon}\times
\frac{1}{3 (\epsilon -1) \epsilon
    (\epsilon +1)^2 (2 \epsilon -3)}
\nonumber\\&&~~~~~~~
\times
\Bigl[
48 (3-2 \epsilon ) C_A \left\{(\epsilon +1)^2 \left(2 \epsilon ^3-9
   \epsilon ^2+8 \epsilon -2\right) n_l-2 (\epsilon -1)^2 \left(2 \epsilon ^4-4
   \epsilon ^3+11 \epsilon ^2-3 \epsilon -2\right) C_A\right\}
   I_{{SS}}^d 
\nonumber\\&&~~~~~~~~~~
   +24 \epsilon ^2
   (\epsilon +1) C_A \left\{2 (7 \epsilon -8) (\epsilon -1)^2
   C_A+(\epsilon +1) n_l\right\} I_{{SS}}^e
\nonumber\\&&~~~~~~~~~~
   +8 \Bigl\{ (\epsilon +1) C_A
   \Bigl(\left(408 \epsilon ^6-1633 \epsilon ^5+2479 \epsilon ^4-1818 \epsilon ^3+698
   \epsilon ^2-149 \epsilon +15\right) C_F
\nonumber\\&&~~~~~~~~~~~~~
   -3 \left(18 \epsilon ^5-42 \epsilon ^4+13
   \epsilon ^3+34 \epsilon ^2-32 \epsilon +7\right) n_l\Bigr)
   +6 \left(-3 \epsilon ^3+\epsilon ^2+3 \epsilon -1\right)^2 C_F
   n_l
\nonumber\\&&~~~~~~~~~~~~~
   -(\epsilon -1)^2
   \left(456 \epsilon ^5-815 \epsilon ^4-96 \epsilon ^3+387 \epsilon ^2+184 \epsilon
   -108\right) C_A^2\Bigr\} \,I_{{SS}}^g 
\nonumber\\&&~~~~~~~~~~
   - 4^{\epsilon } \pi ^{\epsilon -2} e^{-\gamma_E  \epsilon }
   (\epsilon -1) (\epsilon +1)^2 (2 \epsilon -3) \left\{2 (\epsilon -1) C_A+(1-2 \epsilon)
    C_F\right\} \left(11 C_A-2 n_l\right) {iI_S^b}
\Bigr]
\,,
\label{PotGeneDim}
\eea
where $\gamma_E=0.5772\dots$ denotes the Euler constant, and
the one-loop and two-loop master integrals are given by
\bea
&&
iI_S^b=
\frac{16^{\epsilon -1} \pi ^{\epsilon } }{\cos (\pi  \epsilon )\Gamma
   (1-\epsilon )}
\,, 
\\&&
I_{{SS}}^d = 
 \frac{16^{\epsilon -2} \pi ^{2 \epsilon -4} \Gamma \left(\frac{3}{2}-2 \epsilon \right)^2 \Gamma (1-\epsilon ) \Gamma \left(\epsilon -\frac{1}{2}\right) \Gamma \left(2 \epsilon -\frac{1}{2}\right)}{\Gamma (3-4 \epsilon )}
\,,
\\&&
I_{{SS}}^e = 
 -\frac{8^{2 \epsilon -3} \pi ^{2 \epsilon -2} \csc (\pi  \epsilon ) \Gamma \left(\frac{1}{2}-\epsilon \right)^2 \Gamma \left(\epsilon +\frac{1}{2}\right)}{\Gamma (1-2 \epsilon ) \Gamma \left(\frac{3}{2}-\epsilon \right)}
\,,
\\&&
I_{{SS}}^g = 
\frac{ 2^{6 \epsilon -8} \pi ^{2 \epsilon -2} \csc (\pi  \epsilon ) \Gamma \left(\frac{3}{2}-2 \epsilon \right) \Gamma \left(\frac{1}{2}-\epsilon \right) \Gamma \left(2 \epsilon -\frac{1}{2}\right)}{\Gamma (2-3 \epsilon ) \Gamma \left(\frac{3}{2}-\epsilon \right) \Gamma (\epsilon )}
\,.
\eea
The term proportional to ${iI_S^b}$ originates from the counter-term contribution corresponding
to the renormalization of the gauge coupling constant.

\section*{Acknowledgement}
The author thanks Y.~Kiyo for a suggestion to look into the 
$O(\epsilon)$ term of the $1/(mr^2)$ potential.
The author is also grateful to A.~Penin and M.~Steinhauser for explaining the details of their calculations
of the potential.
The two-loop Hamiltonian, on which the current result is based, was calculated
in collaboration with G.~Mishima and H.~Takaura.
This work was supported in part
by JSPS KAKENHI Grant Number JP23K03404.

\end{document}